\documentclass[10pt,letterpaper]{article}
\usepackage[top=0.85in,left=2.75in,footskip=0.75in]{geometry}

\usepackage{amsmath,amssymb}
\usepackage{bbm}

\usepackage{changepage}

\usepackage[utf8x]{inputenc}

\usepackage{textcomp,marvosym}

\usepackage{cite}

\usepackage{nameref,hyperref}

\usepackage[right]{lineno}

\usepackage{microtype}
\DisableLigatures[f]{encoding = *, family = * }

\usepackage[table]{xcolor}

\usepackage{array}

\newcolumntype{+}{!{\vrule width 2pt}}

\newlength\savedwidth

\raggedright
\usepackage[aboveskip=1pt,labelfont=bf,labelsep=period,justification=raggedright,singlelinecheck=off]{caption}

\makeatletter
\renewcommand{\@biblabel}[1]{\quad#1.}
\makeatother

\usepackage{lastpage,fancyhdr,graphicx}
\usepackage{epstopdf}
\fancyheadoffset[L]{2.25in}
\fancyfootoffset[L]{2.25in}
\newcommand{\vect}[1]{\boldsymbol{{#1}}}

\begin{document}
\vspace*{0.2in}

\begin{flushleft}
{\Large
\textbf\newline{Actin reorganization throughout the cell cycle mediated by motor proteins} 
}
\newline
\\
Maria-Veronica Ciocanel\textsuperscript{1},
Aravind Chandrasekaran\textsuperscript{2},
Carli Mager\textsuperscript{3},
Qin Ni\textsuperscript{4},
Garegin Papoian\textsuperscript{5},
Adriana Dawes\textsuperscript{6}
\\
\bigskip
\textbf{1} Department of Mathematics and Biology, Duke University, Durham, NC, USA
\\
\textbf{2} Department of Mechanical and Aerospace Engineering, University of California San Diego, La Jolla, CA, USA
\\
\textbf{3} Department of Biochemistry, The Ohio State University, Columbus, OH, USA
\\
\textbf{4} Department of Chemical and Biomolecular Engineering, University of Maryland, College Park, MD
\\
\textbf{5} Department of Chemistry and Biochemistry and Institute for Physical Science and Technology, University of Maryland, College Park, MD, USA
\\
\textbf{6} Department of Mathematics and Department of Molecular Genetics, The Ohio State University, Columbus, OH, USA
\bigskip

%
%





* ciocanel@math.duke.edu

\end{flushleft}

\section*{Abstract}
Cortical actin networks are highly dynamic and play critical roles in shaping the mechanical properties of cells. The actin cytoskeleton undergoes significant reorganization over the course of the cell cycle, when cortical actin transitions between open patched meshworks, homogeneous distributions, and aligned bundles. Several types of myosin motor proteins, characterized by different kinetic parameters, have been involved in this reorganization of actin filaments. Given the limitations in studying the interactions of actin with myosin \textit{in vivo}, we propose stochastic agent-based model simulations and develop a set of data analysis measures to assess how myosin motor proteins mediate various actin organizations. In particular, we identify individual motor parameters, such as motor binding rate and step size, that generate actin networks with different levels of contractility and different patterns of myosin motor localization. In simulations where two motor populations with distinct kinetic parameters interact with the same actin network, we find that motors may act in a complementary way, by tuning the actin network organization, or in an antagonistic way, where one motor emerges as dominant. This modeling and data analysis framework also uncovers parameter regimes where spatial segregation between motor populations is achieved. By allowing for changes in kinetic rates during the actin-myosin dynamic simulations, our work suggests that certain actin-myosin organizations may require additional regulation beyond mediation by motor proteins in order to reconfigure the cytoskeleton network on experimentally-observed timescales.
  
\section*{Author summary}
Cell shape is dictated by a scaffolding network called the cytoskeleton. Actin filaments, a main component of the cytoskeleton, are found predominantly at the periphery of the cell, where they organize into different patterns as the cell progresses through the cell cycle. The actin filament reorganizations are mediated by motor proteins from the myosin superfamily. Using a stochastic model to simulate actin filament and motor protein dynamics and interactions, we systematically vary motor protein kinetics and investigate their effect on actin filament organization. Using novel measures of spatial organization, we quantify conditions under which motor proteins, either alone or in combination, can produce the different actin filament organizations associated with progression during the cell cycle. These results yield new insights into the role of motor proteins, as well as into how multiple types of motors can work collectively, to produce specific network patterns through the cell cycle.

\section*{Introduction}
Virtually all cells contain a cytoskeleton, a collection of structural filaments that are required for critical processes including division and migration \cite{Hohmann2019}.
The cytoskeleton consists of three major classes of filaments: actin filaments, microtubules, and intermediate filaments.
The actin cortex, a thin meshwork of actin filaments just below the cell membrane, is a major constituent of the cytoskeleton.
Indeed, actin accounts for 10\% or more of a cell's total protein, making it one of the most abundant proteins \cite{Pollard2016}.
Actin filaments are highly dynamic, growing and shrinking through the gain and loss of individual actin monomers.
Actin filaments are also polar, with distinct polymerization kinetics at the two ends resulting in directionally biased filament growth.
The end that favours actin monomer addition is called the barbed (or plus) end, while the end that is less favourable for polymerization is called the pointed (or minus) end.
Cells rely on the dynamic nature of actin filaments to respond quickly to internal and external cues by reorganizing the actin cortex.
These actin reorganizations can result in shape changes and variations in the mechanical properties of the cell \cite{Schaks2019,Pollard2017,Pollard2019,Taneja2020}.
Despite our understanding of the complex filament-level dynamics of actin, conditions that favor formation of specific actin network architectures are still poorly understood, and thus the formation of cortical actin networks is the focus of this study.

\begin{figure}[!ht]
    \centering
    \includegraphics[width=0.8\textwidth]{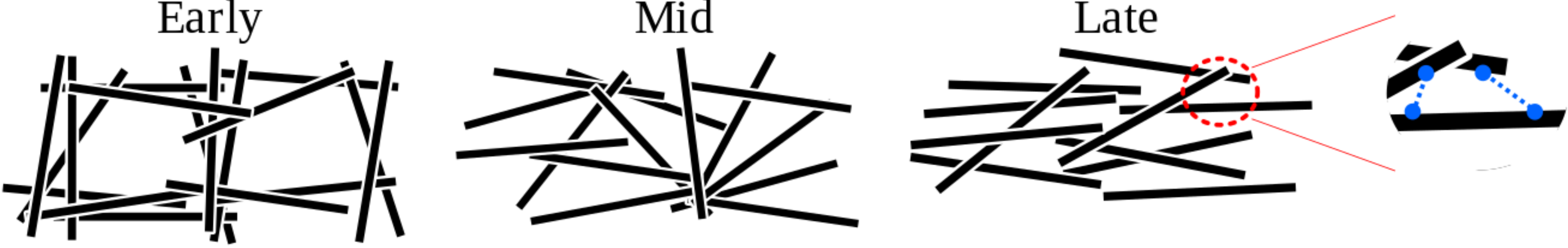}
    \caption{{\bf Typical actin reorganization in one patch of the cortex over the cell cycle.}
    Early stages in the cell cycle are characterized by an open meshwork of actin filaments (black), which transitions to clusters, or foci, distributed throughout the cortex. Later in the cell cycle, actin filaments tend to be more aligned and oriented towards the cytokinetic furrow. Detail of a potion of the meshwork (right) shows filaments crosslinked by proteins such as the motor protein myosin (blue).
    }
    \label{fig:actin_organization}
\end{figure}

Cortical actin reorganization is particularly striking over the course of the cell cycle (Figure \ref{fig:actin_organization}).
For instance, in the early embryo of the nematode worm \textit{Caenorhabditis elegans (C. elegans)} \cite{Strome1988,Hill1988}, cortical actin filaments are initially organized in an open meshwork, characterized by patches with few filaments.
This open meshwork is reconfigured into a homogeneous, isotropic mesh.
As the early embryo prepares for first division, cortical actin filaments are aligned at the middle of the embryo to form the cytokinetic ring, with filaments outside the cytokinetic ring orienting towards the division plane.
These actin reorganizations occur over the course of approximately 15 minutes in the early embryo.
This substantial reorganization of cortical actin through the cell cycle is a common theme in many cell types, from plants \cite{panteris1992organization,traas1987actin,yu2006visualization} to mammals \cite{jackson1989relationship,jalal2019actin}.

In addition to polymerization dynamics, actin filaments are transported by the activity of motor proteins, particularly those from the myosin superfamily.
Members of the myosin superfamily bind to actin filaments and hydrolyze ATP during their power stroke to generate force, resulting in movement of individual filaments. 
In many cells, including the early \textit{C. elegans} embryo, Type II myosins, also called conventional myosins, are implicated in reorganization of cortical actin filaments.
The family of Type II myosins, which consists of the proteins NMY-1 and NMY-2 in \textit{C. elegans} \cite{Piekny2003}, assemble into mini-filaments, with multiple heads containing actin binding domains at either end of the mini-filament.
This bipolar mini-filament structure allows Type II myosins to simultaneously bind two actin filaments, moving the bound filaments relative to each other.
Type II myosins are minus end directed motor proteins, meaning that they take a ``step'' towards the minus end of the actin filament during their power stroke, resulting in movement of the actin filament in the direction of its plus end. 
These myosins are also non-processive, meaning they release from the actin filament after a single power stroke and do not continue to ``walk'' along the actin filament.
The released myosin diffuses until it finds another pair of actin filaments available for binding.
Myosin can also be prevented from performing a power stroke if the force applied to a bound myosin mini-filament is greater than its stall force.
Type II myosins are critical for the cortical reorganization observed over the cell cycle \cite{Houdusse2016,Ideses2013,Dalous2008}.

Eukaryotic cells contain approximately 40 different myosin genes \cite{Spudich2010}, and many of these other myosin motor proteins are thought to be involved in actin cytoskeleton organization during the cell cycle. For instance, myosin V is an unconventional myosin which transports cargo as it moves along actin filaments \cite{Trybus2008}, and plays a critical role in fission yeast cytokinesis \cite{Laplante2015,Wang2014}.
Myosin VI, which moves towards the minus end of an actin filament \cite{Spudich2010}, segregates to distinct spatial locations throughout the cell cycle \cite{Majewski2012}, and is critical for cell proliferation in certain cancers \cite{Zhang2016}.
While the dynamics of these other myosin motors are less well understood compared to Type II myosins, it is clear that kinetic parameters associated with these different myosins can vary widely. For instance, non-muscle myosin IIA motors are thought to have an individual head step size of $6$~nm during the power stroke \cite{vilfan2003instabilities,popov2016medyan}, whereas the myosin V motor has been found to have a mean step size of $36$~nm \cite{clemen2005force}, and the processive myosin VI motor has a broader distribution of step sizes, with mean forward steps of $\sim 30$~nm \cite{rock2001myosin}. Similarly, the characteristic unbinding force of the non-muscle myosin IIA motor is assumed to be $12.6$~pN in \cite{erdmann2013stochastic,popov2016medyan}, while muscle myosin II has been found to have a measured average unbinding force of $9.2$~pN \cite{mikhailenko2010insights}, and the myosin-V unbinding force has been estimated as 3--5~pN \cite{oguchi2008load}. 
Given the wide variability in kinetic parameters associated with myosin motor proteins, and a lack of comprehensive information about the dynamics and role of different motor proteins in regulating actin cytoskeleton organization through the cell cycle, we focus here on the effect of different parameters associated with motor protein activity.
In particular, we quantify changes in actin cytoskeleton organization as a result of variation in individual kinetic rates.
In this way, we characterize changes in the actin cytoskeleton observed over the cell cycle without constraining the simulations to a specific motor protein and its properties.

A number of mathematical models have been proposed to investigate the formation of higher order actin structures due to the activity of myosin.
Continuum models consisting of PDEs have shown the ability of motor proteins with different characteristics to reproduce experimentally observed structures \cite{White2014,White2015,Cytrynbaum2006}.
This approach allows for analysis of the corresponding model, but does not take into account the noisy interactions of individual filaments or motors. In addition, these continuum mathematical models are unable to capture the structural evolution of the interacting proteins at the molecular level. 
Stochastic models that explicitly simulate individual filament and motor dynamics have been used to yield insights into the dynamics of actin and myosin structures. A review and comparison of existing agent-based cytoskeletal models is provided in \cite[Table~1]{popov2016medyan}. While the different models vary in their implementation, these frameworks consistently show that changes in motor protein activity can induce different actin organizations \cite{chandrasekaran2019remarkable,Belmonte2017}.
These modeling approaches have yielded insights into the range of actin-based structures that can be formed in the presence of motor proteins such as Type II myosins, but is it not yet understood how motor proteins can efficiently and robustly transition between the actin organization observed through the cell cycle or how they might coordinate to establish the observed actomyosin structures.

In this investigation, we use the stochastic simulation platform MEDYAN to simulate the organization and transition between different actin organizations.We adapt data analysis measures to characterize the simulated actin structures and quantify the time course of their formation. We find that a single motor protein is capable of producing the actin structures associated with cell cycle progression, with variations in motor protein step size, binding rates, stall force, and number of motor heads resulting in the greatest changes in actomyosin organization.
In particular, changes in these parameters can produce actin structures associated with different stages of the cell cycle, ranging from tightly clustered foci to loose meshworks of filaments. 
When two motor protein populations with different kinetic parameters interact with the same actin meshwork, additional properties emerge: actin structures may adopt an intermediate organization, between the two extremes of the motor proteins acting alone; one motor protein may dominate, entirely dictating the structure of the actin meshwork with the second motor protein acting as passive cargo; and we also observe some motor protein segregation, with motor proteins occupying distinct spatial regions.
Additionally, we find that transitioning between between actin structures can be achieved in a more timely manner when two motor proteins act together. Together, these results demonstrate the importance of cooperation between motor proteins to efficiently construct and reorganize the actin structures associated with cell cycle progression.

\section{Stochastic simulation framework for actin-myosin interactions}\label{sec:medyan}

We carry out mechanochemical simulations of actin-myosin interactions using the MEDYAN (Mechanochemical
Dynamics of Active Networks) modeling framework developed in \cite{popov2016medyan}. This simulation package uses a coarse-grained representation of interacting semi-flexible polymers (actin filaments) in three dimensions. The cytoskeletal network mechanics are integrated with stochastic reaction-diffusion processes, whose dynamics are calculated using the \textit{next reaction method} \cite{gibson2000efficient}. The simulation space is divided into compartments and diffusing molecules are assumed to be uniformly mixed within each compartment. Stochastic movement between compartments is used to model the diffusion and molecular transport of various chemical species. 

Here we use MEDYAN to model actin filament polymerization phenomena, in addition to essential crosslinker ($\alpha$-actinin) processes such as binding and unbinding. Additional active processes involving motor protein (such as myosin II minifilament) binding, unbinding, and walking are incorporated in the model. The force fields employed to model the actin filaments, as well as their interaction potentials with linkers and motors that characterize filament deformations, are detailed in \cite{popov2016medyan}. To understand actin-myosin organization in the simulation domain, we take advantage of the mechanical modeling of actin filaments, which consist of cylinder units with equilibrium spacing \cite{chandrasekaran2019remarkable}. Further details about the MEDYAN model framework and implementation can be found in \cite{popov2016medyan,komianos2018stochastic,chandrasekaran2019remarkable,ni2019turnover}. 

In our simulations, we consider a $2 \mu m \times 2\mu m \times 0.2 \mu m$ domain, with cubical compartments with side length of $0.2 \mu m$. As in previous work \cite{ciocanel2019topological}, we  carry out standard implementations of the model in \cite{popov2016medyan}, which is parameterized for actin filament polymerization and depolymerization, $\alpha$-actinin cross-linking proteins, and non-muscle myosin IIa motor filaments; however, we use larger numbers of myosin motors, consistent with the myosin concentrations used in computational studies of actin bundles \cite{chandrasekaran2019remarkable}. We are especially interested in understanding motor regulation and how it impacts cytoskeleton organization. In particular, we wish to investigate whether variation in parameters associated with motor protein activity can account for the diversity and dynamics of actin-based structures observed over the course of the cell cycle.

In addition, we quantify the emergent actin-myosin organization in simulations where two populations of motors with distinct properties interact with actin filaments.
This is motivated by studies into the maintenance of ring channels, circular openings in developing \textit{C. elegans} oocytes, which suggest that the Type II myosins NMY-1 and NMY-2 act antagonistically to maintain a stable ring channel opening.
Further, it was shown that NMY-1 and NMY-2 occupy spatially distinct regions near the ring channel opening \cite{Coffman2016}. Thus, we wish to investigate the conditions and kinetic parameters under which motor proteins may segregate into spatially distinct regions.

\section{Results}

\subsection{Myosin motor parameters influence the emerging actin organization}\label{1motor}

We begin by considering the dynamic organization of actomyosin networks in the presence of one myosin motor population. 
In this manuscript, we refer to bipolar aggregates of myosin molecules as myosin motors. Such bipolar minifilaments stay bound longer and can work more efficiently compared to individual myosin molecules. To accurately model minifilaments based on the implicit properties of individual myosins, we use the parallel cluster model, which offers a rigorous statistical mechanics-based paradigm to understand the emergent behaviors of a group of myosins \cite{erdmann2013stochastic}. In this model, the kinetic parameters of binding, walking, stalling, and unbinding of myosins can be predicted based on individual myosin properties such as binding rate, unbinding rate, stall force, and unbinding force. In addition, we can also account for the variability in the number of myosins in a population of minifilaments.

In \cite{popov2016medyan}, motor parameters are chosen to model the behavior of non-muscle myosin IIA minifilaments. Since various myosin motors have been hypothesized to exert force on the actin cytoskeleton throughout the cell cycle, we investigate the impact of different motor properties on cytoskeleton organization. We build on the simulation framework and baseline parameter values for non-muscle myosin IIA motors in \cite{popov2016medyan} (see Table~\ref{tab:baseline}) to uncover such differences in actomyosin organization. By changing one motor parameter at a time and characterizing the resulting organization using the data analysis methods described in \S~\ref{sec:methods}, we suggest potential mechanisms of motor regulation that may be responsible for changes in actin assembly throughout the cell cycle. Since MEDYAN simulations are stochastic, the dynamic actomyosin organization may vary across simulations; unless otherwise noted, we consider ten independent stochastic runs for each parameter setting.

\begin{table}[!ht]
\begin{adjustwidth}{-2.25in}{0in}
\caption{{\bf Baseline parameters for the myosin II minifilaments from \cite{popov2016medyan}.}}
\label{tab:baseline}
\center
\begin{tabular}{lllll}
Motor & Meaning & Value & Reference & Range\\
parameter & &  & & \\
\hline \hline
  $d_\mathrm{step}$ & Motor step size  &    $6$~nm & \cite{vilfan2003instabilities,popov2016medyan} & 3--36 \\
  $k_\mathrm{head,bind}$ & Per-head binding rate  &    $0.2$/s  & \cite{kovacs2003functional,popov2016medyan}  & 0.1--0.8\\
  $F_s$ & Stall force of motor minifilament  &    $100$~pN  & \cite{popov2016medyan} & 10--200\\
  $\mathrm{range}_\mathrm{heads}$ & Range for number of heads of the minifilament  &    15--30 &  \cite{verkhovsky1993non,popov2016medyan} & 2--45\\
  $K_\mathrm{motor}$ & Motor stretching force constant  &    $2.5$~pN  & \cite{vilfan2003instabilities,popov2016medyan}  & 1.25--10 \\
  $F_0$ & Per-head unbinding force  &    $12.62$~pN  & \cite{erdmann2013stochastic,popov2016medyan} & 6--25\\
  $\mathrm{range}_{\mathrm{rxn}}$ & Range of motor binding reaction  &    175--225~nm  & \cite{popov2016medyan} & 125--275
\end{tabular}
\end{adjustwidth}
\end{table}

\subsubsection{Baseline parameter simulations}
We begin by describing the actomyosin organization under the baseline parameter values in Table~\ref{tab:baseline} using the data analysis methods in \S~\ref{sec:methods}. In these baseline simulations, the myosin motors dynamically organize the actin cytoskeleton into 1-2 clusters with some stray filaments; Figure~\ref{fig:baseline_snapshots}A illustrates several time snapshots of a sample actin-myosin simulation (see Supplementary Video S1 for complete sample simulation). The radial distribution function described in \S~\ref{method_rad_dist} is shown in Figure~\ref{fig:baseline_snapshots}B for the same simulation times, averaged over ten stochastic MEDYAN runs. This illustrates that the inter-monomer distance distribution changes from a wide peak at medium distance (radius) values to a large peak at small values corresponding to the filaments that are clustering together, as well as a flatter peak at larger distances corresponding to actin cylinders positioned in different clusters. 

Additional methods of characterizing the actin cytoskeleton organization are provided in Figure~\ref{fig:baseline_time_series}A. The actomyosin network radius of gyration (described in \S~\ref{method_Rg} and introduced for this system in \cite{popov2016medyan}) shows that the different runs exhibit a range of behaviors, with some simulations leading to a small increase in the radius of gyration (decreased network contractility) and others leading to a small decrease in the radius of gyration (increased network contractility). Since there is no global alignment of filaments in the simulation domain, the orientational order parameter (described in \S~\ref{method_globalalign} and introduced for this system in \cite{popov2016medyan}) does not show significant changes through time. Finally, the actin organization shows an overall clustered distribution in the spatial statistics measure described in \S~\ref{method_spatstat} (as opposed to a regular or spatially random distribution of actin cylinders). 

\begin{figure}[!ht]
\centering
\includegraphics[width=0.9\textwidth]{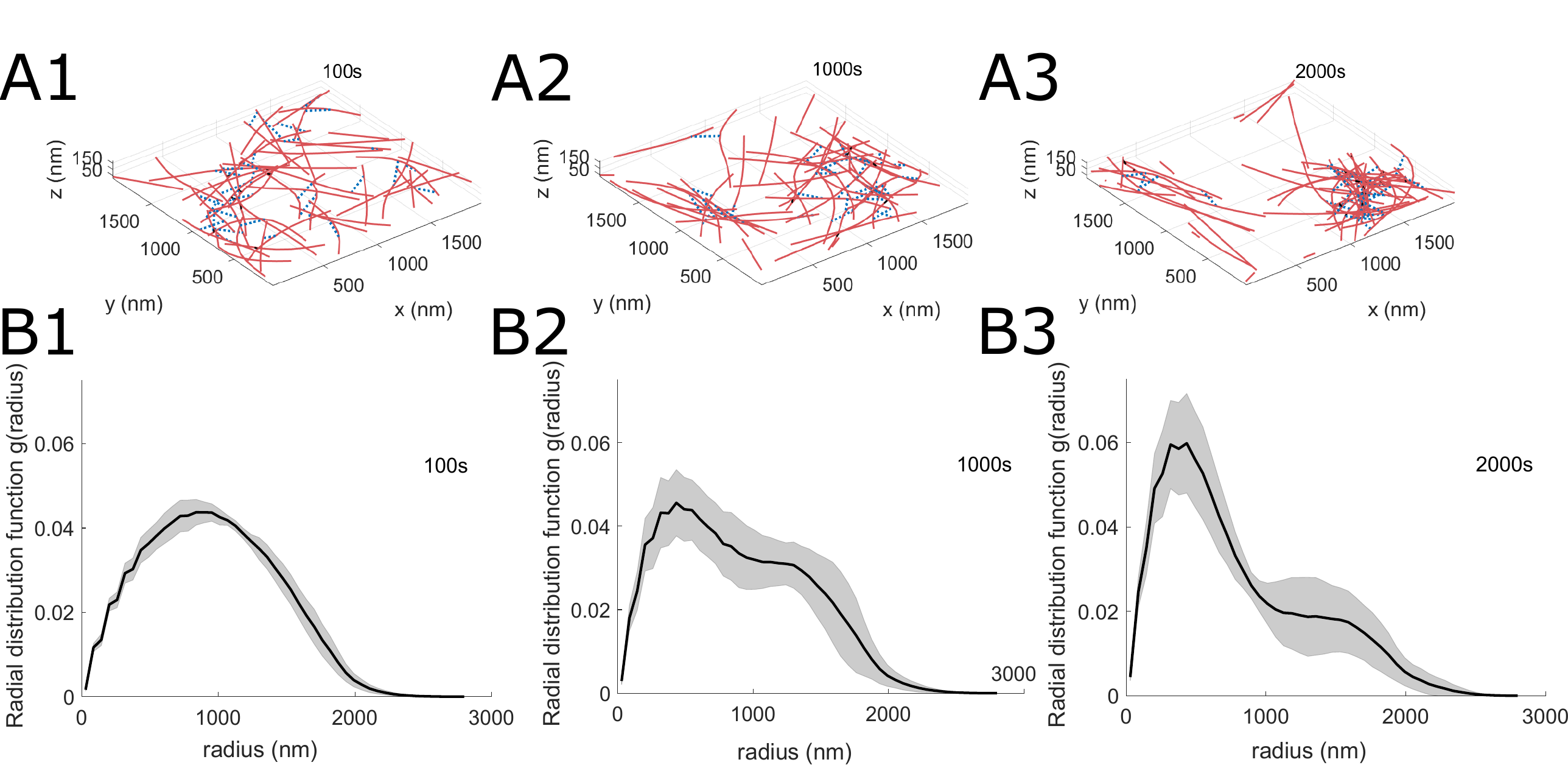}
\caption{{\bf Simulation results using standard parameters (Table~\ref{tab:baseline}).}
Baseline simulation snapshots at times (A1) $t$=100, (A2) $t$=1000, and (A3) $t$=2000s, with actin filaments depicted as long red polymers, myosin motors represented as medium-length dashed blue lines and cross-linkers shown as short black lines. Parameter values used are given in Table \ref{tab:baseline}. (B1-B3) Radial distribution function, indicating the density of pairwise distances between actin filaments (\S~\ref{method_rad_dist}), for the corresponding single time point snapshots in (A). The radial distribution function indicates filaments become more clustered over time, consistent with the simulation snapshots.
In all panels, solid lines indicate the average, and shaded areas indicate the standard deviation over 10 independent stochastic runs.
See Figure \ref{fig:baseline_time_series} for further measures of actin filament and motor protein organization in these simulations.}
\label{fig:baseline_snapshots}
\end{figure}

The myosin motor organization is visualized in Figure~\ref{fig:baseline_time_series}B using a three-dimensional motor localization plot as a function of time and of distance from each motor to the center of the domain (further described in \S~\ref{method_motor_loc}); this average radial motor localization with respect to the middle of the domain does not change significantly through time. However, the measure defined in \S~\ref{method_motorconvhull}, which calculates the area of the boundary polygon around the myosin motors, shows a steady decline through time, indicating that the motors are overall localizing in space as they cluster actin filaments into tighter actomyosin structures. This is further confirmed using the spatial statistics-based measure described in \S~\ref{method_spatstat}, which increases through time and therefore suggests that the distribution of the motor protein pattern becomes increasingly more clustered through time. 

\begin{figure}[!ht]
\centering
\includegraphics[width=0.9\textwidth]{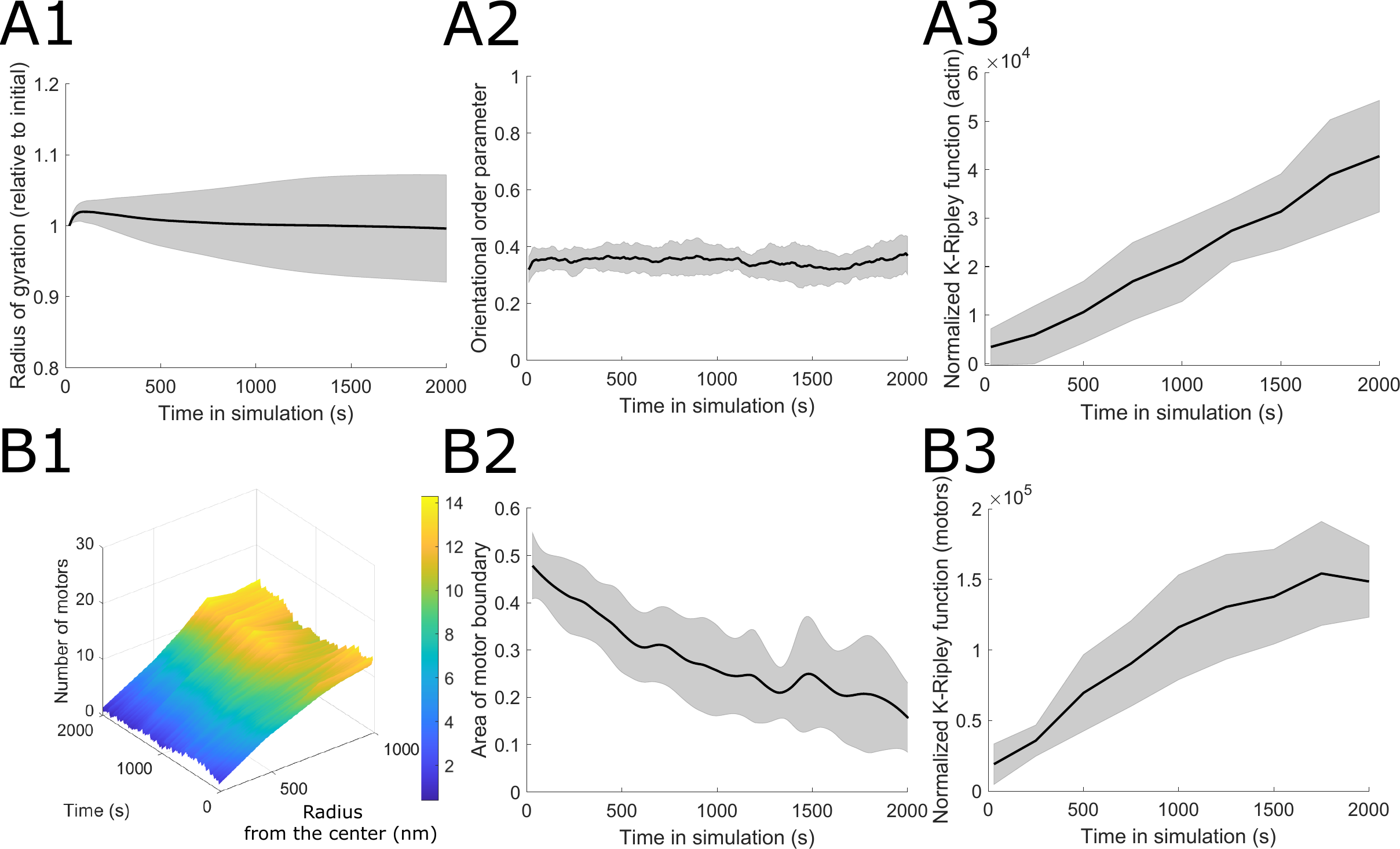}
\caption{{\bf Additional measures to characterize (A) actin filament and (B) motor protein organization in simulations using baseline parameter values (Table \ref{tab:baseline}).}
(A1) The radius of gyration, a measure of filament network contractility (\S~\ref{method_Rg}), shows little change over time but with high variability around the mean.
(A2) Similarly, the orientational order, indicating the degree to which actin filaments are aligned over the whole domain (\S~\ref{method_globalalign}), shows little change over time with the baseline parameter values.
(A3) The normalized K-Ripley function, a spatial statistic measure which indicates the extent of uniformity in the spatial distribution of filaments (\S~\ref{method_spatstat}), increases over time reflecting the increasingly clustered filament organization.
(B1) Myosin motor distribution is visualized over time as a function of the distribution of radial distances from the center of the domain (\S~\ref{method_motor_loc}), indicating a higher density of motors at the periphery of the domain.
(B2) The area of motor boundary, determined by the minimum polygon that encloses all motors on the domain (\S~\ref{method_motorconvhull}), decreases over time reflecting the increased clustering of both filaments and motors.
(B3) As in panel (A3), the normalized K-Ripley function for the myosin motors increases over time as a result of increased filament and motor clustering. 
Solid lines indicate the average and shaded areas indicate the standard deviation over 10 independent stochastic runs.
See Figure~\ref{fig:baseline_snapshots} for simulation snapshots and further measures of actin filament organization.}
\label{fig:baseline_time_series}
\end{figure}

In the following sections, we present variations in motor parameters that lead to significant changes in cytoskeleton organization as compared to the baseline. In Table~\ref{tab:parameters}, we summarize how these parameters affect microscale aspects of myosin minifilament behavior in the MEDYAN model as well as how they impact network-level filament and motor organization. 

\begin{table}[!ht]
\begin{adjustwidth}{-2.25in}{0in}
\caption{{\bf Impact of the parameters discussed in \S~\ref{subsec:1motor_stepsize}, \ref{subsec:1motor_onrate}, \ref{subsec:1motor_stallforce}, and \ref{subsec:1motor_nrheads} on myosin minifilament behavior (MEDYAN model in \cite{popov2016medyan}) and on network-level cytoskeleton organization (this study).}}
\label{tab:parameters}
\center
\begin{tabular}{lllll}
Parameter & Myosin minifilament behavior changed & Impact of motor parameter increase\\ & & on cytoskeleton network behavior \\
\hline \hline
 $d_\mathrm{step}$ & Base walking rate  &  Tighter and faster cluster formation  \\
  $k_\mathrm{head,bind}$ & Base walking rate, filament unbinding  &  Cluster formation initially, \\ & & loose network for increasingly large values \\
  $F_s$ & Minifilament walking rate  &  Increase in clustering and contractility  \\
  $\mathrm{range}_\mathrm{heads}$ & Filament unbinding  &    Increase in clustering and contractility
\end{tabular}
\end{adjustwidth}
\end{table}

\subsubsection{Step size}\label{subsec:1motor_stepsize}
\begin{figure}[!ht]
\centering
\includegraphics[width=0.9\textwidth]{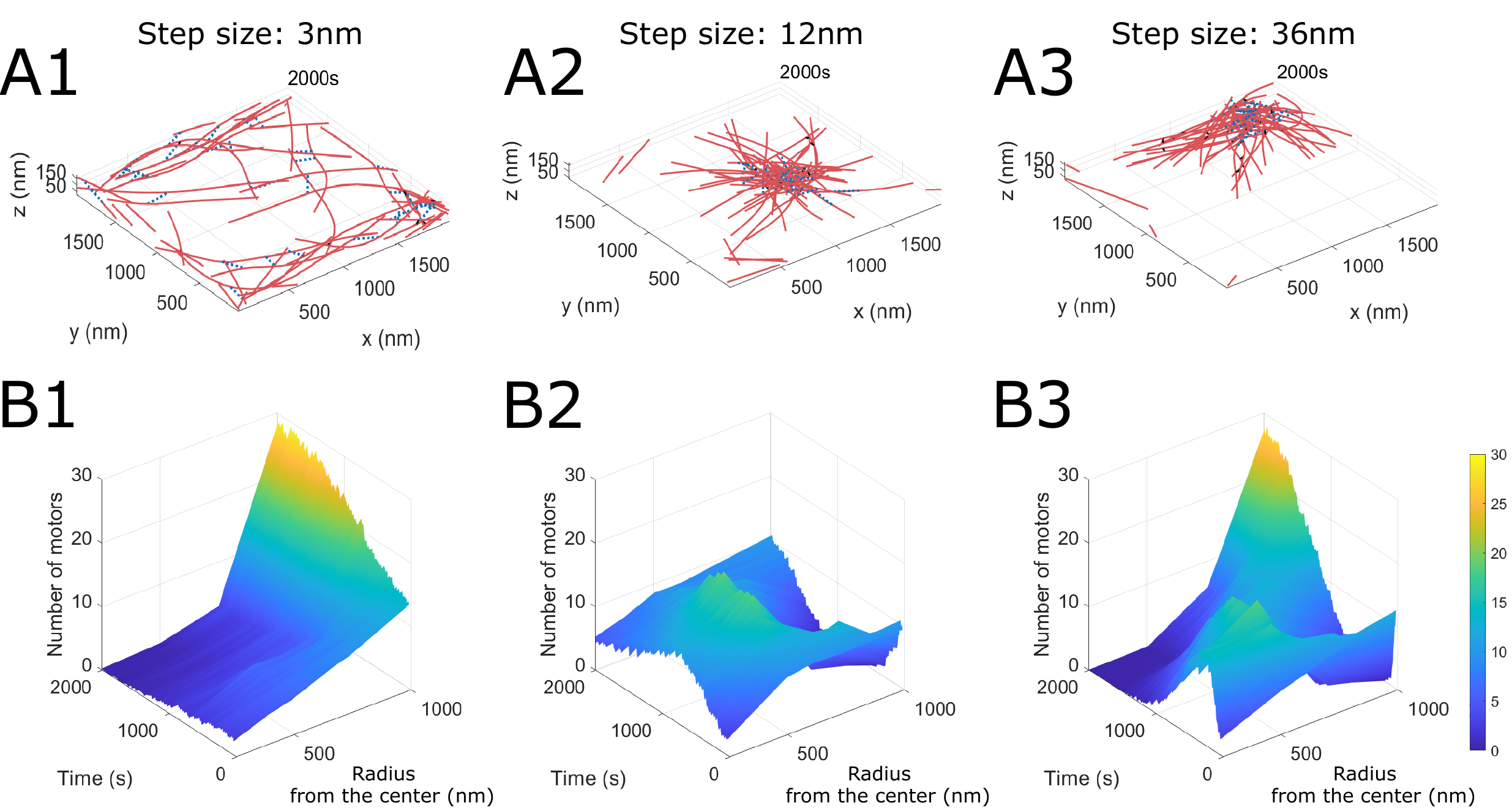}
\caption{{\bf Increasing motor step size leads to actin filament clustering. }
(A) Final simulation snapshots ($t$=2000s) at the indicated motor step size ((A1): 3nm, (A2): 12nm, (A3): 36nm), showing aggregation of actin filaments for larger step sizes.
(B) The corresponding motor localization over the whole domain for the different motor step sizes. 
}
\label{fig:stepsize_each}
\end{figure}

We refer to the physiological binding distance of a single myosin motor head $d_\mathrm{step}$ as the motor step size. In the MEDYAN model, this parameter affects the base walking rate of the motors: $k^0_\mathrm{fil,walk}=\frac{d_\mathrm{step}}{d_\mathrm{total}} \frac{1-\rho}{\rho}k_{\mathrm{head,bind}}$, where $\rho$ is the motor duty ratio, $d_\mathrm{total}$ is the distance between binding sites on the model actin cylinders, and $k_\mathrm{head,bind}$ is the single head binding rate \cite{popov2016medyan}. Figure~\ref{fig:stepsize_each}A shows the actin-myosin organization at the final time of sample simulations with small ($3$~nm) and large ($12$ and $36$~nm) myosin step sizes relative to the baseline value. The small step size leads to considerably more spread out filament organization, with some filament alignment at the domain boundaries, consistent with the motor localization in Figure~\ref{fig:stepsize_each}B. On the other hand, the larger step sizes lead to more compact contractile actin-myosin clusters, with motors localized in these clusters. Increasing the step size leads to an increase in the base motor walking rate, so that myosin motors have better access to filaments and therefore lead to their contraction.

\begin{figure}[!ht]
\centering
\includegraphics[width=0.6\textwidth]{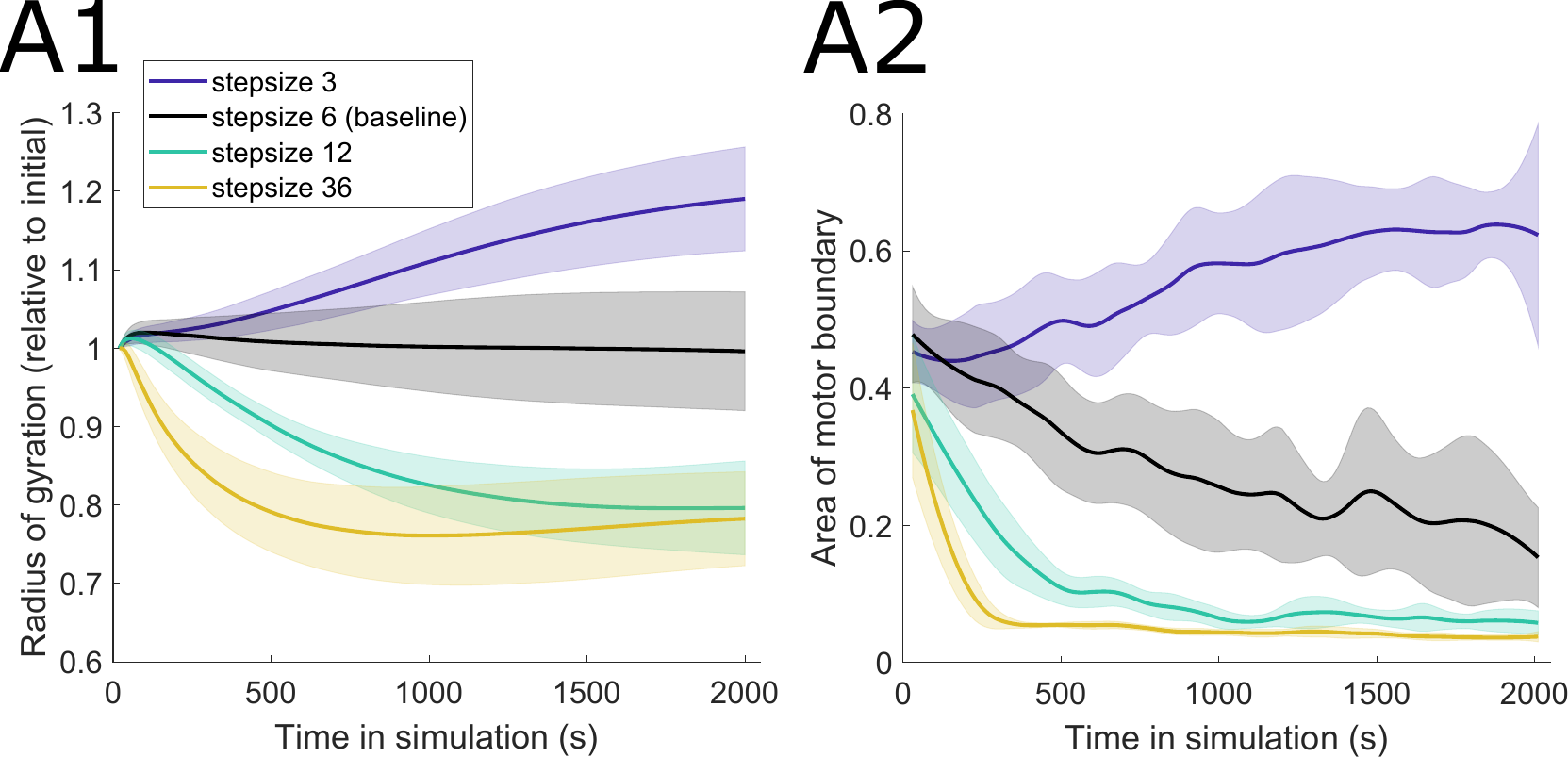}
\caption{
{\bf Actin and myosin organization is consistent across simulations for varying myosin step sizes. }
(A1) The radius of gyration decreases as motor step size is increased, reflecting the formation of actin clusters (Figure~\ref{fig:stepsize_each}A).
(A2) The area of the polygon enclosing the motor proteins also decreases as motor step size increases, due to motor proteins binding to clustered actin filaments.
Solid lines indicate the average and shaded areas indicate the standard deviation over 10 stochastic runs.
See Figure~\ref{fig:baseline_time_series} for additional measure details.}
\label{fig:stepsize_time_series}
\end{figure}

This behavior of the actin-myosin organization is similar across additional stochastic runs, as illustrated by the time-series measures in Figure~\ref{fig:stepsize_time_series}. The actomyosin network radius of gyration and the area of the motor boundary both increase at small step sizes, reflecting the relaxing of the filaments into a more homogeneous distribution for small step sizes. For larger step sizes, the radius of gyration and the motor area decrease through time, showing faster establishment of contractile clusters. 

\subsubsection{Binding rate}\label{subsec:1motor_onrate}
\begin{figure}[!ht]
\centering
\includegraphics[width=0.9\textwidth]{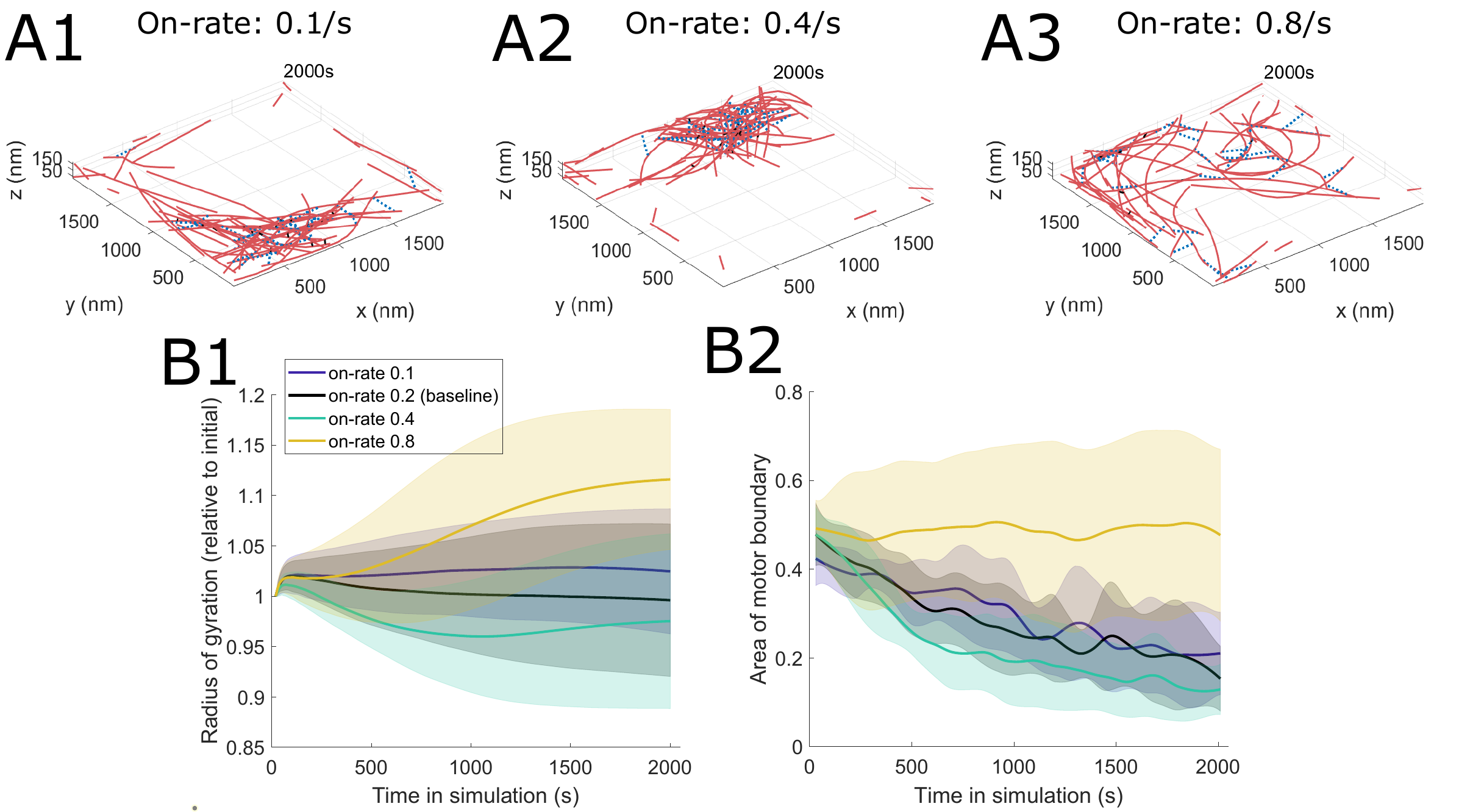}
\caption{
{\bf Motor protein binding on-rate has a nonlinear effect on actomyosin organization, with high and low rates resulting in a loose meshwork.}
(A) Final simulation snapshots ($t=2000$s) for different myosin binding on-rates ((A1): 0.1/s, (A2): 0.4/s, (A3): 0.8/s). (B) Characterizing the actin and myosin time-series organization in simulations with varying myosin on-rates. (B1) The radius of gyration reflects the nonlinear impact of binding on-rate, with intermediate values resulting in a more clustered organization. (B2) Similarly, the polygon boundary is smallest for intermediate binding on-rate values. Solid lines indicate the average and shaded areas indicate the standard deviation over 10 stochastic runs indicating high variability between different runs. See Figure~\ref{fig:baseline_time_series} for additional measure details.}
\label{fig:onrate_time_series}
\end{figure}

We denote the per-head motor binding rate by the on-rate $k_\mathrm{head,bind}$. Increasing this parameter leads to an increase in the base walking rate of the motors, but also affects the base filament unbinding rate in a nonlinear way according to the parallel cluster model for non-processive motors \cite{erdmann2013stochastic} used in MEDYAN \cite{popov2016medyan}. The small binding rate sample simulation in Figure~\ref{fig:onrate_time_series}A shows a slightly more spread out cytoskeleton organization, whereas the large binding rate simulation ($0.4$/s) displays compact clusters with fewer free filaments than in the baseline case. In general, increasing the on-rate leads to an increase in the motor's duty ratio (the proportion of time that a head spends in the bound state) and therefore yields a larger number of bound heads, so that myosin motors reside on the filaments longer and contract them. However, further increasing this rate to $0.8$/s leads to considerably looser and more spread out actomyosin organization, with no noticeable clustering  (see Supplementary Video S2). In this case, the motors reside on the filaments much longer and appear less mobile, so that they play more of a cross-linking role in the dynamic actin organization.   

These observed behaviors are consistent across simulations, as illustrated by the time-series measures in Figure~\ref{fig:onrate_time_series}B. Although there is more variability across model runs for the $0.8$/s binding rate, the actin organization relaxes into a more homogeneous distribution in this parameter setting, while the myosin motors spread out across a larger portion of the domain. We note that, unlike the linear change in actomyosin behavior as a result of varying the step size $d_\mathrm{step}$ in \S~\ref{subsec:1motor_stepsize}, the system behaves nonlinearly as the head binding rate $k_\mathrm{head,bind}$ increases; this is due to the fact that the latter parameter can impact multiple mechanisms in the model (base walking rate, base filament unbinding rate), thus providing additional and more nuanced insights on the impact of myosin motor parameters on the cytoskeleton organization.

\subsubsection{Stall force}\label{subsec:1motor_stallforce}
\begin{figure}[!ht]
\centering
\includegraphics[width=0.9\textwidth]{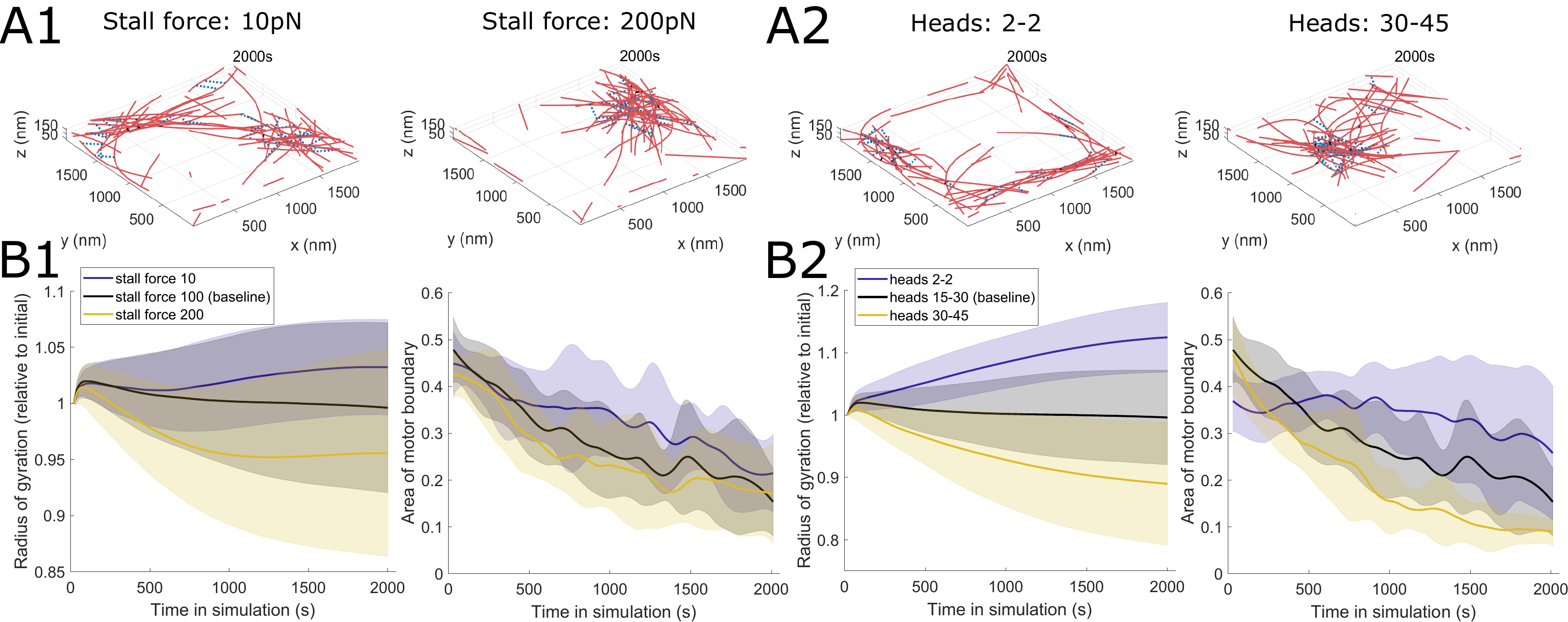}
\caption{{\bf Increasing motor protein stall force and number of heads per minifilament results in clustered actin filaments.} (A) Final simulation snapshots ($t=2000$s) for different myosin stall forces ((A1): 10pN and 200pN) and different ranges of numbers of heads ((A2): minimum of 2 to maximum of 2 (2-2), and minimum 30 to maximum 45 (30-45)). (B) Characterizing the actin and myosin time-series organization through radius of gyration and spatial distribution of motor proteins in simulations with varying myosin stall forces (B1) and different ranges of numbers of heads (B2). Both measures decrease as the stall force or number of heads increases, visually consistent with the increased clustering in the simulations. Solid lines indicate the average and shaded areas indicate the standard deviation over 10 stochastic runs. See Figure~\ref{fig:baseline_time_series} for additional measure details.}
\label{fig:force_heads_all}
\end{figure}

We let $F_s$ denote the stall force of a myosin motor minifilament. In MEDYAN, this parameter impacts the motor minifilament walking rate: $k_\mathrm{fil,walk}(F_\mathrm{ext})=\max\{0,k_\mathrm{fil,walk}^0\frac{F_s-F_\mathrm{ext}}{F_s+F_\mathrm{ext}/\alpha}\}$, where $k_\mathrm{fil,walk}^0$ is the base walking rate of the motors, $F_\mathrm{ext}$ is the external force or tension experienced by the myosin filament, and $\alpha$ is a parameter that tunes the strength of the dependence on the external force \cite{popov2016medyan}. Figure~\ref{fig:force_heads_all}A1 shows that increasing stall force is associated with an increase in clustering and contractility, since the walking rate stays larger for higher external forces experienced by the motor. While there is more variability in the stochastic runs associated with this parameter,  Figure~\ref{fig:force_heads_all}B1 is consistent with this observation that larger stall forces lead to a more contractile actin network and to slightly tighter spatial segregation of the motors. 

\subsubsection{Number of motor heads}\label{subsec:1motor_nrheads}
We let $N_t$ represent the number of heads in the myosin filament, and vary the range of this parameter. $N_t$ influences the base filament unbinding rate according to the parallel cluster model for non-processive motors \cite{erdmann2013stochastic} used in MEDYAN \cite{popov2016medyan}, and is directly proportional to the zero force residence time. Figure~\ref{fig:force_heads_all}A2 shows that allowing for exactly $2$ heads of the myosin minifilament (a dimer configuration) leads to a more spread out actin organization, with alignment at the domain boundaries, while larger numbers of heads (minimum $30$ and maximum $45$) yield more compact cluster organization. This observation is also summarized using the measures characterizing the dynamic actin and myosin organization in Figure~\ref{fig:force_heads_all}B2. Increasing the number of heads $N_t$ leads to a decrease in the base filament unbinding rate, so that myosin motors are less likely to unbind from actin and therefore consistently create actomyosin clusters.

\subsubsection{Open meshwork organization}
While some of the parameter settings investigated above illustrate an opening in the actin organization and some filament alignment at the boundaries, another means of generating an open actin-myosin meshwork is to reduce the number of myosin motor minifilaments in the MEDYAN model simulations. Supplementary Video S3 shows the progression to a more open meshwork in the simulation domain with a decrease in the motor number as well as with a reduction in the motor stepsize. This suggests that regulation of the availability of the active motors may be one way to generate the open cytoskeletal meshworks observed in early stages of the cell cycle.

\subsection{Two-motor populations contribute to tuning of cytoskeletal organization and reveal dominant motors}\label{2motor}

Experimental observations have shown that several types of myosin motors \cite{Coffman2016} or multiple populations of the same myosin motor with characteristics that depend on the local cellular environment \cite{east2011regulation} may regulate actomyosin organization during the cell cycle. Motivated by these observations, we study the impact of two-motor populations with different motor parameters on simulations of cytoskeletal networks in MEDYAN. We consider the same total number of motors as in \S~\ref{1motor}, divided into equal numbers of motors for each of the two motor populations of interest.

\subsubsection{Tuning behavior of motor populations with different parameters}
\begin{figure}[!ht]
\centering
\includegraphics[width=0.9\textwidth]{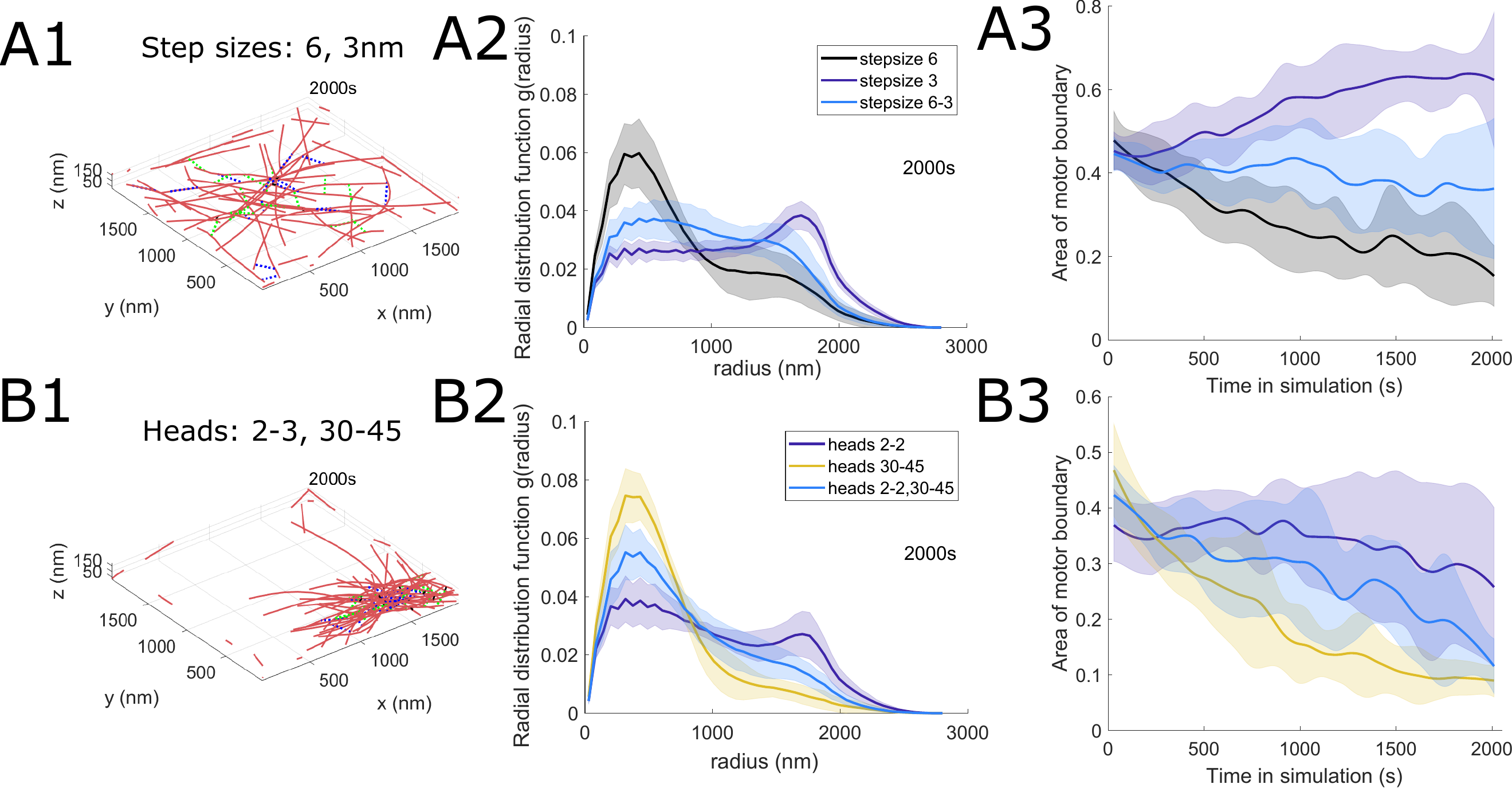}
\caption{{\bf Two populations of motor proteins with different motor step sizes and motor heads result in actin filament organization that lies between the values expected from the motors acting individually.}
Sample final simulation snapshots ($t=2000$s; A1 and B1) and evidence of compromise in actomyosin network behavior in simulations with two motor populations with (A) different myosin motor step sizes (6nm and 3nm), and (B) different ranges for the numbers of myosin motor heads (min 2 to max 2 (2-2), and min 30 to max 45 (30-45)). Measures of radial distribution (A2 and B2) and area of the polygon bounding all motor proteins (A3 and B3) fall between the values calculated for the motor proteins acting alone (compare to Figures~\ref{fig:stepsize_time_series} and \ref{fig:force_heads_all}A2, B2). Solid lines indicate the average and shaded areas indicate the standard deviation over 10 stochastic runs.}
\label{fig:tuning_2motors}
\end{figure}

In many of the model simulations performed, we find that the actomyosin organization is tuned so that the measures of cytoskeletal network behavior lie in-between those corresponding to the behaviors of interactions with a single motor population (i.e., characterized by a single motor parameter set). Two sample examples for interactions with motors with $3$ vs. $6$~nm step sizes as well as with dimers (exactly $2$ heads at each end of the minifilament) vs. motors with $30-45$ heads are shown in Figure~\ref{fig:tuning_2motors}; as in the case of simulations with one motor population, we represent actin filaments as red polymers and cross-linkers as short black lines, while here myosin motors with the first parameter value are shown in dashed blue lines and with the second parameter value, in green dashed lines. In both examples, the distribution of actin inter-monomer distance distribution at the final simulation time is balanced between the distributions resulting from simulations with each of the single motor populations. Similarly, the measure that quantifies myosin motor localization with time (calculated for all motors in the simulation) reflects the same observation that actomyosin behavior is tuned compared to the single-motor population settings.

\subsubsection{Dominant behavior of certain motors}
\begin{figure}[!ht]
\centering
\includegraphics[width=0.9\textwidth]{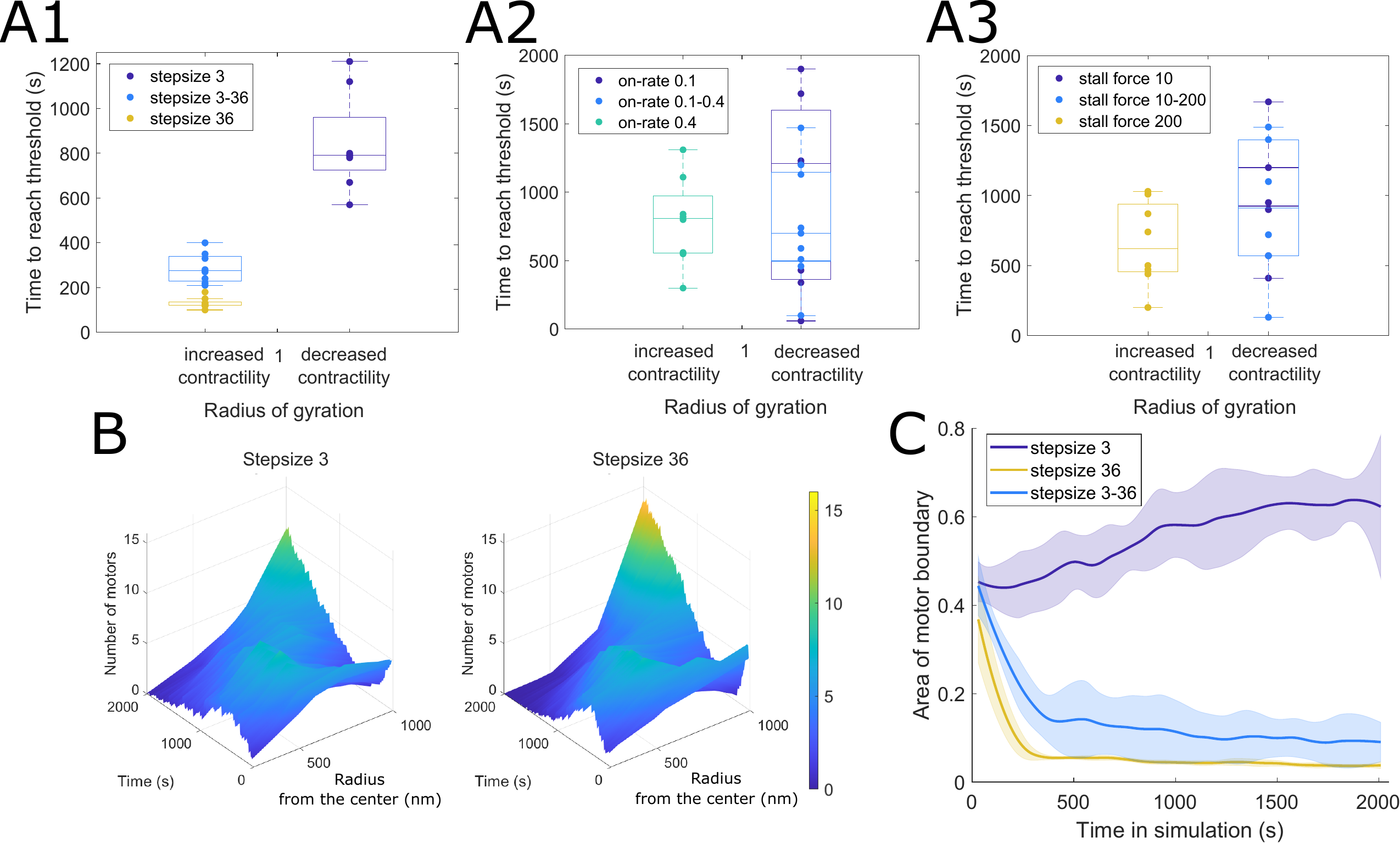}
\caption{{\bf In two motor population simulations, motors with large step sizes, low binding rates, and low stall forces dominate actomyosin organization dynamics.}
(A) Timing to reach threshold in contractility ($\pm 10\%$ in left panel, $\pm 5\%$ in center and right panels), showing that (A1) large  step size, (A2) low binding rates, and (A3) low stall forces dominate the dynamics of contractility. Compare to Figures~\ref{fig:stepsize_time_series}, \ref{fig:onrate_time_series}, \ref{fig:force_heads_all}. (B) Localization of motors and (C) Motor spread in simulations with motors with $3$~nm and $36$~nm step sizes, further demonstrating that the large step size dictates the dynamics of the ensemble independently of the dynamics contributed by the motor protein with a smaller step size.  Solid lines indicate the average and shaded areas indicate the standard deviation over 10 independent stochastic runs.}
\label{fig:dominant_2motors}
\end{figure}

In certain two-motor population simulations, we find that one of the motors dominates the dynamics and is able to re-position the other motor population. In these parameter settings, the dominant motors may dictate the overall actin organization. To illustrate this, we build on the network contractility measure in \S~\ref{method_Rg} to determine the first time in each simulation when the radius of gyration increases or decreases by a certain threshold amount (determined based on the relative increase or decrease in contractility observed for that parameter). The box plots in Figure~\ref{fig:dominant_2motors}A show several examples where one motor population (with a specific step size, binding rate, and stall force) dominates the actin network organization in interactions with another motor population; see Figure~\ref{fig:baseline_time_series}A for baseline simulations for an example where the radius of gyration stays on average around the normalized value of $1$. Figure~\ref{fig:dominant_2motors}B,C further focuses on model interactions of myosin motors with $3$~nm and $36$~nm step sizes with actin. Both motor populations localize similarly throughout time and space as shown in Figure~\ref{fig:dominant_2motors}B, and the behavior resembles the localization plot for myosin motors with $36$~nm step size only in \ref{fig:stepsize_each}B. This is also reflected by the area of the motor boundary polygon measure in Figure~\ref{fig:dominant_2motors}C, overall suggesting that the $3$~nm motor population is passively transported and organized by the dominant $36$~nm motor population.

\subsubsection{Motor segregation in certain parameter regimes}

\begin{figure}[!ht]
\centering
\includegraphics[width=0.9\textwidth]{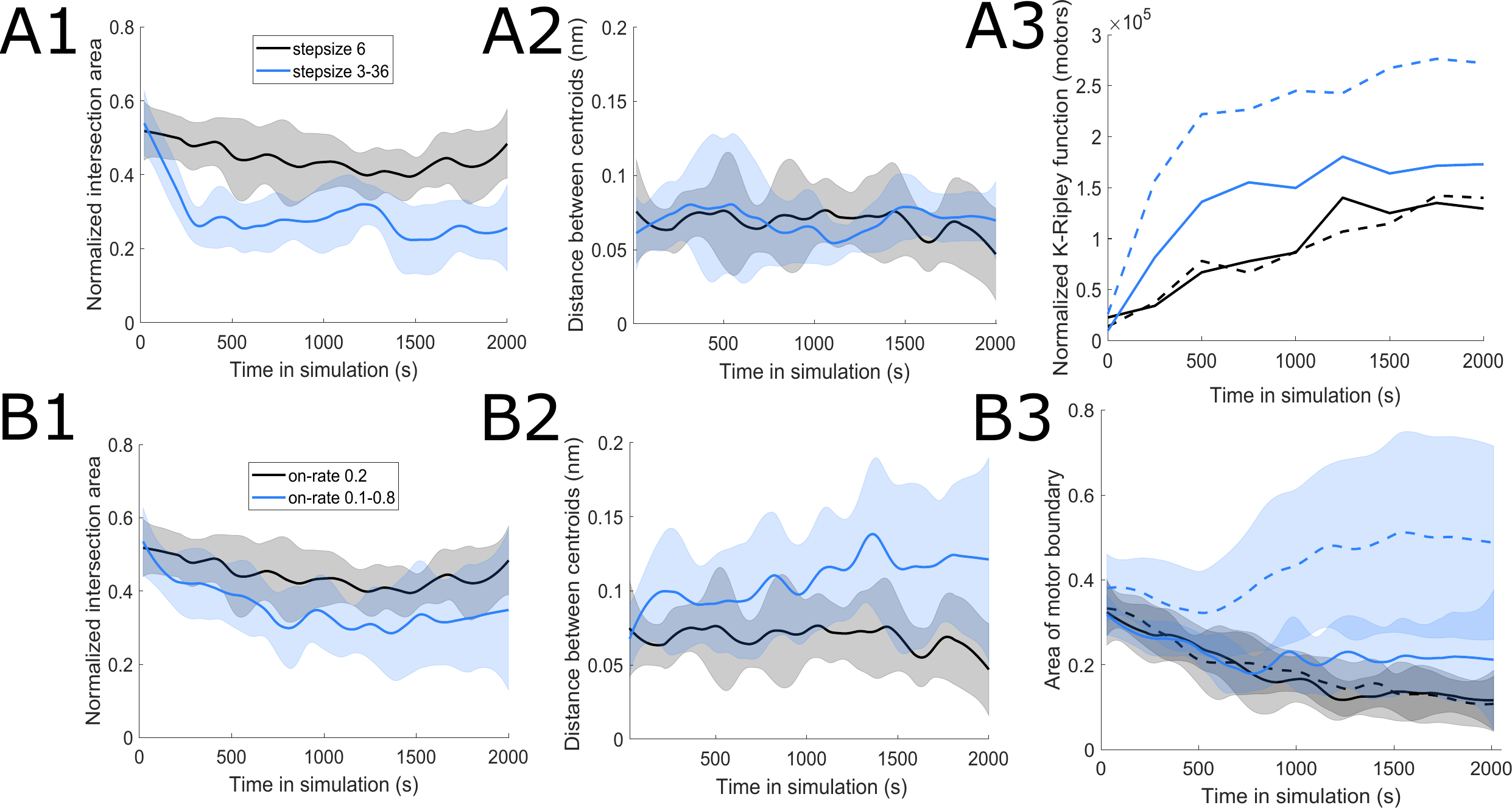}
\caption{{\bf Two motor populations with (A) different step sizes or (B) different binding rates self-organize into spatially distinct domain.} (A1) and (B1) shows decreasing intersection area of the two motor protein populations (blue line) compared to two identical motor populations with baseline parameters (black line; Figure~\ref{fig:baseline_time_series}).
(A2) and (B2) show displacement of the centroid of the polygon bounding the motor proteins, further suggesting spatial segregation. (A3) The normalized K-Ripley function, which measures the spatial distribution of each motor protein population (\S~\ref{method_spatstat}), and (B3) the area of one of the motor population's boundary both increase over time, suggesting the actin filaments become more clustered; the first motor population is indicated with blue solid lines ($3$~nm step size in A, and $0.1$/s on-rate in B), and the second motor population with dashed blue lines ($36$~nm step size in A, and $0.8$/s on-rate in B). The black curves correspond to simulations with two motor populations with identical baseline parameters. Solid and dashed lines indicate the average and shaded areas indicate the standard deviation over 10 independent stochastic runs.}
\label{fig:segregation_2motors}
\end{figure}

In few of the parameter regimes investigated, we found evidence of some spatial segregation of the two motor populations interacting with actin filaments. In Figure~\ref{fig:segregation_2motors}, we use the measures described in \S~\ref{motor_segreg} and \ref{method_spatstat} to analyze the interactions of actin filaments with motors with $3$ vs. $36$~nm step sizes as well as with motors with on-rates of $0.1$/s vs. $0.8$/s. The measures described in \S~\ref{motor_segreg} rely on finding the two-dimensional boundary polygons around the point clouds consisting of each motor's center locations; we then compute the intersection area between these boundary polygons for the two motor species as well as the distance between the centroids corresponding to the two polygons. Figure~\ref{fig:segregation_2motors}A shows that the intersection area of the boundaries for motors with the two different step sizes decreases in time compared to baseline simulations, thus suggesting that the motors might segregate in space; however, the distance between the centroids of the motor boundaries does not change significantly. This is because actomyosin is consistently organized in a tight cluster for this motor combination, with motors in both categories localizing closer together through time. To further clarify the distribution pattern of the two motors, we apply the spatial statistics measure described in \S~\ref{method_spatstat} to each motor population. The right panel of Figure~\ref{fig:segregation_2motors}A shows that the $36$~nm step size motor has an even tighter cluster distribution within this actomyosin network. Figure~\ref{fig:segregation_2motors}B suggests that there is distinct spatial segregation of the motors with $0.1$ vs. $0.8$/s binding rates, given that the normalized intersection area of their boundaries decreases and the centroid distance between these motor boundaries increases through time. We further confirm this by visualizing the spread measure for each motor population in the right panel: the small on-rate motor forms a cluster through time, whereas the large on-rate motor is distributed throughout the simulation domain given its less mobile behavior (as also observed in \S~\ref{subsec:1motor_onrate}).

\subsection{Transitions in motor parameters reflect the remodeling ability of the cytoskeleton}\label{time_transition}

To understand the capacity of the actomyosin network to re-organize under myosin motor regulation during the cell cycle or in cells where local ATP abundance is altered, we implement a MEDYAN framework where the myosin motor binding rate can change at a specified time point during the simulation. In particular, we consider the setting where the myosin binding rate switches between $0.8$/s and $0.4$/s. We chose these parameter values for our study since, as shown in Figure~\ref{fig:onrate_time_series} in \S~\ref{subsec:1motor_onrate}, the $0.8$/s binding rate leads to loose actomyosin organization whereas the $0.4$/s binding rate generates organizations with compact clusters. 

\begin{figure}[!ht]
\centering
\includegraphics[width=0.8\textwidth]{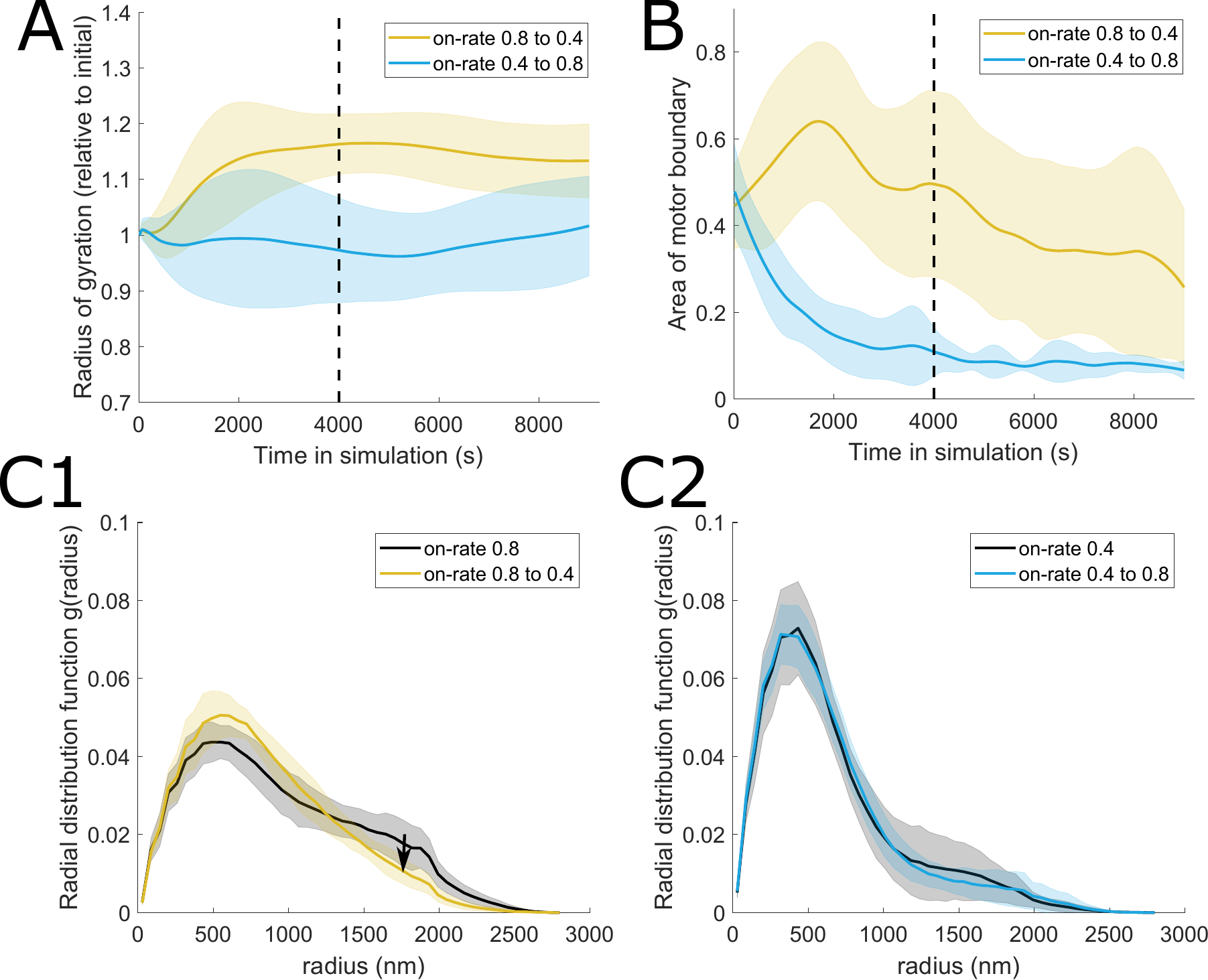}
\caption{{\bf Time-dependent change in motor protein binding rate results in asymetric changes in actin filament organization.} Changing the myosin motor binding rate at $4000$~s on (A) actomyosin network contractility, (B) motor spread, and (C) radial distribution function results in actin filament organization that depends on the order in which the motor proteins interact with the actin meshwork. The dashed vertical lines in A-B indicate the time when the on-rate parameter changes in the simulation. Both panels of (C) indicate the distribution of monomer distances at $4000$~s in black and the distribution of distances after the parameter change, at $9000$~s, in the colors consistent with panels A-B. Solid lines indicate the average and shaded areas indicate the standard deviation over 10 independent stochastic runs.}
\label{fig:time_change_onrate}
\end{figure}

In Figure~\ref{fig:time_change_onrate}, we explore changes in this motor parameter $4000$~s into simulations that last a total of $9000$~s, in order to allow the system to equilibrate. 
In \S~\ref{subsec:1motor_onrate}, we found that a relatively large binding rate ($0.8$/s) results in a spread out actin organization, with less mobile motors given their long residence time on the filaments. Figure~\ref{fig:time_change_onrate}A,B shows that the average actin contractility measure undergoes a very slow decrease following the switch to motor binding rate $0.4$/s and that the myosin motors slowly become more localized in the simulation domain  (see Supplementary Video S4 for a sample simulation). Similarly, this parameter change leads to a slightly less homogeneous distribution as illustrated by the better contouring of one peak (corresponding to a more compact cluster) in the radial distribution function in panel C1 of Figure~\ref{fig:time_change_onrate}. Switching from binding rate $0.4$/s to $0.8$/s shows a slow relaxation from the network's contractile behavior (Figure~\ref{fig:time_change_onrate}A) but no considerable change in the myosin organization (Figure~\ref{fig:time_change_onrate}B) or in the pairwise distances between actin cylindrical segments (Figure~\ref{fig:time_change_onrate}C). This may suggest that a significant change from a contractile actin-myosin network organization may require regulation from additional cell cycle processes.

\section*{Discussion}
Cortical actin undergoes dynamic reorganization throughout the cell cycle in many organisms, adopting a wide variety of configurations from homogeneous meshes to spatially localized clusters. Experimental observations show that many members of the myosin motor family (such as myosin II, V, and VI) may be involved in cytoskeleton remodeling. Even within one myosin family, small differences in motor activity between isoforms can result in different organization of the actin cortex \cite{Taneja2020}. Motivated by these observations, we use stochastic agent-based modeling simulations to investigate how different kinetic parameters describing motor protein activity affect the organization of the actin cytoskeleton, and how multiple motor populations may interact with the same actin meshwork. We propose data analysis measures that assess both the dynamic remodeling of the actin network as well as the clustering and segregation of the myosin motor proteins. While motivated by understanding the interactions of certain myosin motors that have been found to collectively organize the cytoskeleton, we do not constrain our model to specific motor proteins and their corresponding properties, but rather aim to understand how regulation of individual motor kinetic rates can lead to diverse actin network structures.  

Overall, we find that cytoskeleton organization is highly sensitive to the kinetics of interacting motor proteins.
Here we focus on the role of several key kinetic parameters (Table \ref{tab:baseline}): binding rate, stall force, motor step size, and the number of heads per minifilament.
We also studied the influence of motor parameters such as the stretching force constant, the per-head unbinding force, and the reaction range of the motor binding reaction on actin organization; we find that the ranges considered for these parameters (see Table~\ref{tab:baseline}) do not yield significantly different actomyosin organization from the baseline.

When acting individually, a single type of motor protein can produce a range of actin organizations when interacting with an ensemble of actin filaments.
Tight clusters of actin filaments with a smaller radius of gyration are associated with large motor step sizes, higher stall force, and higher number of heads. 
Homogeneous networks, where the actin filaments are loosely spread on the domain, are associated with lower values of those parameters.
Variations in the per-head motor binding rate suggest an optimal intermediate value is needed to organize the actin filaments into tight clusters, with low and high values of this parameter resulting in a loose network. 

When two distinct populations of motor proteins interact with the actin filaments, unexpected behaviours can emerge.
Many motor combinations appear to compromise, for instance when the motors have different step sizes or motor heads, with network measures taking on values that lie between the values when the motors act alone. This suggests that some motor protein properties are complementary and can work together to produce novel organizations of actin filaments.
In contrast, some combinations of motor protein populations appear to act antagonistically, with one motor protein dominating the organization of the meshwork. For instance, motor proteins with a large step size dominate motor proteins with a small step size, with network measures in the presence of both motor proteins almost indistinguishable from those when the motor with a large step size acts alone.
In this case, the less dominant motor acts as a passive cargo, being transported to particular regions in the domain by the activity of the dominant motor; this also means that the passive motor is only able to find actin filaments to bind to based on the cytoskeleton organization imposed by the dominant motor.
This relationship between the motors can be advantageous if the less dominant motor requires a particular filament configuration or spatial localization but cannot achieve the required dynamics when acting alone.
Further, in the presence of two motor protein populations, spatial segregation can be achieved, where motors that are initially homogeneously distributed on the domain self organize into distinct regions.
This type of motor protein sorting has been previously observed \cite{Coffman2016}, and appears to rely on differences in step size and binding rates in our study. Spatial exclusion of filament-bound cytoskeletal components has been an active area of research. Computational studies on spatial segregation of crosslinker populations have shown that effective separation is dependent on actin polymerization rate and mechanical properties such as the size difference between two crosslinkers and the bending modulus of actin filaments  \cite{freedman2019mechanical}. Here, we show that kinetic parameter differences are sufficient to achieve spatial segregation in actin networks with heterogeneous motor populations.

Dynamic transitions in actin network organization can be realized when motor protein kinetics are changed throughout the cell cycle. These kinetic rate changes could be due to inactivation or degradation of one type of motor protein with simultaneous activation or synthesis of another motor protein with different kinetics, or regulatory proteins hiding or exposing functional sites on the motor protein. We find that, when activated in the presence of a particular organization of actin filaments, some motor proteins are unable to remodel the existing actin meshwork, while other motor proteins are able to reconfigure the actin filaments into new structures.

By rigorously quantifying actin filament organization in response to motor protein kinetics, we have demonstrated a range of possible actin-based structures.
Acting individually, motor proteins can produce many of these structures, but efficient transitions between structures require motor proteins to act together.
Changes in motor protein kinetics and cooperation between motor protein types may account for the large scale changes in actin filament organization observed over the cell cycle.

\section{Data Analysis Methods}\label{sec:methods}

\subsection{Network contractility}\label{method_Rg}
To assess the contractile behavior of each actomyosin network generated, we calculate the network radius of gyration \cite{popov2016medyan}. This measure has been shown to be effective in determining the filament contractility in MEDYAN simulations \cite{popov2016medyan}. Each filament in MEDYAN is stored in terms of the locations of the coarse-grained cylinder segments (monomer units) that make up each actin filament. We let $n$ be the total number of cylindrical segments from all actin filaments in one time frame of a MEDYAN simulaiton. Let $\vect{r}_i = (x_i,y_i,z_i)$ be the location of the $i$th cylinder (with coordinates in 3-dimensional space) and determine the geometric center of the ensemble of all cylinders as $\vect{r}_{GC} = ( \mathrm{mean}(x_i),\mathrm{mean}(y_i),\mathrm{mean}(z_i))$. Then the network radius of gyration for that time frame is defined as \cite{popov2016medyan}:
\begin{align}
    R_g &= \sqrt{\frac{1}{n} \sum_{i=1}^n ||\vect{r}_i-\vect{r}_{GC}||^2}\,.
\end{align}
When evaluated at a time-series of MEDYAN simulation frames, the network radius of gyration has a decreasing pattern when the contractile behavior increases through time, and increases when the contractility of the network decreases through time. In visualizations of this contractility measure, we normalize the network radius of gyration $R_G$ by dividing by its value at the first time point in the simulation.

\subsection{Global network alignment}\label{method_globalalign}
To determine the alignment of the actin filaments in the actomyosin network, we calculate an orientational order parameter of the system of actin filaments \cite{popov2016medyan}. This involves setting up the ordering tensor:
\begin{align}
    Q &= \frac{3}{2} \left(\frac{1}{N} \sum_{i=1}^N \vect{\hat{r}_i} \vect{\hat{r}_i}^T - \frac{1}{3} I_3  \right)\,,
\end{align}
and defining the orientational order parameter $S$ as the largest eigenvalue of this tensor; we note that this measure has also been used to assess the alignment of molecules in liquid crystal models \cite{sarman2014director}. Here, $N$ is the number of filaments in the actomyosin network and $I_3$ is the 3 by 3 identity matrix. The normalized direction vector $\vect{\hat{r}_i}$ of each actin filament $i$ is calculated based on the 3-dimensional locations of the two filament ends (the locations of the first and last segments rather than the locations of each cylindrical segment between the ends), which allows for this value to reflect the alignment of the network even when the filaments are bending \cite{popov2016medyan}. If the locations of the ending cylinders are given by $(x_1,y_1,z_1)$ and $(x_2,y_2,z_2)$, then 
$$\vect{\hat{r}_i} = \left(\frac{x_2-x_1}{l},\frac{y_2-y_1}{l},\frac{z_2-z_1}{l}\right)^T\,,$$
with $l$ the length of $\vect{r_i}$. The value $S=0$ of the largest eigenvalue of $Q$ corresponds to random alignment of the filaments in the system, while the value $S=1$ indicates a perfectly aligned filament network \cite{sarman2014director,popov2016medyan}.

\subsection{Radial distribution function}\label{method_rad_dist} We calculate a variation of the radial distribution function to understand the distances between emerging structures in the simulated polymer network. This involves computing the distances between all pairs of actin cylindrical segments in the simulation and binning the distances into a distribution. Letting $\vect{r}_i = (x_i,y_i,z_i)$ be the location of the $i$th cylinder as in \S~\ref{method_Rg}, we determine the matrix of pairwise distances $Z$, where $Z_{ij} = d(\vect{r}_i,\vect{r}_j)$ is the Euclidean distance ($L_2$ norm) between actin monomer unit $i$ and $j$. Noting that the distance between segments ranges from $0$~nm to $2000\sqrt{2}$~nm (the maximum distance if the actin segments are at opposite corners of the domain), we divide this range into $50$ bins and denote the centers of the bins by $\mathrm{radius}_j$. For each time frame $t$, we then define:
\begin{align}
    g(\mathrm{radius}_j,t) &= \frac{1}{N_s(N_s-1)}\sum_{i=1}^{N_s} \sum_{j=1,j\neq i}^{N_s} \mathbbm {1}_{ [\mathrm{radius}_j,\mathrm{radius}_{j+1})} (Z_{ij}) \,.
\end{align}
Here $\mathbbm {1}_A(x)$ is the indicator function with value $1$ when $x \in A$ and $0$ when $x \notin A$. 
$N_s$ is the number of cylindrical segments at time $t$ and therefore the normalization is done by dividing by the number of pairs of actin cylindrical segments.

\subsection{Motor localization}\label{method_motor_loc}
To quantify the spatio-temporal localization of motors in the simulation domain, we divide the simulation domain into cylindrical annuli and determine the number of motors bound to filaments in each such volume and at each time. The thin $z$ dimension of the simulation domain gives the height of each cylinder and the circles centered at $x=y=1000$~nm with radii $0,250,500,750$, and $1000$~nm provide bounds between the annuli. Note that the first volume is actually a cylinder with center at $(1000,1000)$, while the last volume extends outside the boundaries of the cubic simulation domain (this is no concern since motor proteins will simply not be found there). Using the locations of the centers of the minifilaments $m_x,m_y,m_z$, we record the number of myosin motors that are bound to filaments at each time point and count how many are located in each cylindrical annulus.

\subsection{Motor spread} \label{method_motorconvhull}
We aim to quantify the spread of a myosin motor population in MEDYAN actin-myosin simulations. To simplify computation and due to the small height of the domain, we restrict our attention to the centers of the minifilaments in $x$-$y$ space $(m_x,m_y)$. We apply the \texttt{boundary} and \texttt{polyshape} functions in MATLAB to the population of motors, thus generating a 2-dimensional boundary polygon around the motors. We use a default shrink factor of $0.5$ for the \texttt{boundary} method, which means that the resulting polygon around the myosin motors is tighter than the convex hull of the points. Let the polygon around the motor population at time $t$ be denoted by $\mathcal{P}(t)$; then we introduce a measure for the myosin motor spread in the domain:
\begin{align}
    A_{mot}(t) &= \frac{\mathrm{area}(\mathcal{P}(t))}{L^2} \,, 
\end{align}
where $L=2000$~nm is the side of the square domain and thus the polygon area is normalized by the area of the 2-dimensional simulation domain considered.

\subsection{Motor segregation measures} \label{motor_segreg}
Building on the framework for the motor spread measure in \S~\ref{method_motorconvhull}, we introduce two measures for determining the segregation of two motor populations in MEDYAN actin-myosin simulations. We similarly consider the centers of the motor minifilaments in two dimensions $(m_x,m_y)$. Using the \texttt{boundary} and \texttt{polyshape} functions in MATLAB for each population of motors as above, we generate two-dimensional boundary polygons around each motor population. We denote the polygon around the first motor population at time $t$ by $\mathcal{P}_1(t)$ and the one around the second motor population by $\mathcal{P}_2(t)$. Let $(c_{x,1},c_{y,1})$ and $(c_{x,2},c_{y,2})$ be the two-dimensional positions of the centroids of polygons $\mathcal{P}_1(t)$ and $\mathcal{P}_2(t)$. We then define the following measures for the normalized intersection area and the distance between the centroids of the two polygons:
\begin{align}
    A_{int}(t) &= \frac{\mathrm{area}(\mathcal{P}_1(t) \cap \mathcal{P}_2(t))}{\mathrm{area}(\mathcal{P}_1(t) \cup \mathcal{P}_2(t))} \,, \\ 
    D_{cent}(t) &= \frac{\sqrt{(c_{x,1}-c_{x,2})^2 + (c_{y,1}-c_{y,2})^2 }}{L\sqrt{2}}\,,
\end{align}
where $L=2000$~nm is the side of the square domain. The measure $A_{int}$ is normalized against the area of the union of polygons $\mathcal{P}_1(t)$ and  $\mathcal{P}_2(t)$, so as to capture the intersection area relative to the space that both motor populations cover. The measure $D_{cent}$ is similar to the separation distance measure proposed in \cite{ku2010identifying} for the distance between F-actin and myosin-II fluorescence areas.

\subsection{Spatial statistics}\label{method_spatstat}
Spatial statistical methods are useful in understanding the distribution patterns of proteins \cite{paparelli2016analyzing}. Motivated by the use of protein pattern analysis in microscopy images as described in \cite{paparelli2016analyzing}, we use the K-Ripley function to understand how random, cluster, or regular distributions may form in the simulation domain for actin monomer units and myosin motor proteins. As in the previous method, we focus on the locations of proteins in the $x$-$y$ space. For actin, we sample $30\%$ of the monomer units along each filament (as done in \cite{ciocanel2019topological}) to obtain the corresponding point process. For motor proteins, we directly use the locations of the centers of the myosin minifilaments. To analyze these point processes, we calculate the K-Ripley function (using the \texttt{spatstat} function in \textsf{R}), which measures the number of neighbors within a certain radius $r$ to a point \cite{paparelli2016analyzing}:
\begin{align}
    K(r) &= \frac{1}{\lambda} \sum_i \sum_{j \neq i} \frac{\mathbbm{1}_{[0,r)}(d_{ij})}{N} \,,
\end{align}
where $\lambda$ is the density of points in the studied region, $d_{ij}$ is the distance between points $i$ and $j$, and $N$ is the number of points in the dataset. We record the normalized form $H(r)$ of the K-Ripley function:
\begin{align}
    H(r) &= \sqrt{\frac{K(r)}{\pi}} - r 
\end{align}
and note that $H(r)=0$ for complete spatial randomness, $H(r)>0$ for clustering, and $H(r)<0$ for regularity in the distribution of the point process. We therefore record the signed area under the curve of $H(r)$ at nine time points throughout the simulation (every $250$ seconds); larger values for this measure correspond to more clustered patterns in the distribution of actin monomers or of myosin motor proteins.

\section*{Supporting information}

\paragraph*{S1 Video.}
\label{S1_Video}
{\bf Evolution of the cytoskeleton network for baseline parameters.} Sample actin-myosin cytoskeleton organization using MEDYAN for the baseline parameters in Table~\ref{tab:baseline}.

\paragraph*{S2 Video.}
\label{S2_Video}
{\bf Evolution of the cytoskeleton network with varying binding rates.} Sample MEDYAN simulations with binding rates ranging from 0.1 to 0.8 /s show an initial increase in contractility with increasing binding rate, while the larger binding rate generates a more spread out actin organization.

\paragraph*{S3 Video.}
\label{S3_Video}
{\bf Evolution of the cytoskeleton network with varying numbers of motors and step sizes.} Sample MEDYAN simulations with 8 to 32 myosin motors and step sizes ranging from 3 to 36 nm show the progression to a more open actin-myosin meshwork with a decrease in the motor number as well as with a reduction in the motor step size.

\paragraph*{S4 Video.}
\label{S4_Video}
{\bf Evolution of the cytoskeleton network with a time-dependent change on binding rate.} Sample MEDYAN simulation with a change in binding rate from 0.8 to 0.4 /s at 4000s shows a very slow reorganization from a loose into a tighter network.

\section*{Acknowledgments}
M.-V.C. has been supported by The Ohio State University President's Postdoctoral Scholars Program and by the Mathematical Biosciences Institute at The Ohio State University through the National Science Foundation (NSF) DMS-1440386. 
A.T.D. is supported by the National Science Foundation (NSF) under Grant DMS-1554896.

\nolinenumbers

\bibliographystyle{plos2015}
\bibliography{references}

%
%
%


\end{document}